\documentclass[twocolumn,showpacs,preprintnumbers,amsmath,amssymb,showkeys]{revtex4}

\usepackage{ifthen}
\usepackage{ifpdf}
\usepackage{color}

\ifpdf
\usepackage{graphicx}
\usepackage{epstopdf}
\else
\usepackage{graphicx}
\usepackage{epsfig}
\fi

\graphicspath{{/users/physics/dcohen/PREP/}}

\usepackage{latexsym}
\usepackage{amsmath}
\usepackage{amssymb}
\usepackage{bm}
\usepackage{wasysym}

\newcommand{\mass}{\mathsf{m}}

\newcommand{\tbox}[1]{\mbox{\tiny #1}}
 
\newcommand{\be}[1]{\begin{eqnarray}\ifthenelse{#1=-1}{\nonumber}{\ifthenelse{#1=0}{}{\label{e#1}}}}
\newcommand{\beq}{\begin{eqnarray}}
\newcommand{\eeq}{\end{eqnarray}} 

\newcommand{\hide}[1]{\textcolor{red}{[hidden text]}}

\newcommand{\Eq}[1]{\textcolor{blue}{Eq.\!\!~(\ref{#1})}} 
\newcommand{\Fig}[1]{\textcolor{blue}{Fig.}\!\!~\ref{#1}}

\newcommand{\mycite}[1]{\textcolor{blue}{\cite{#1}}}

\begin{document}

\title{Energy absorption by ``sparse" systems: beyond linear response theory}

\author{Doron Cohen}

\affiliation{Departments of Physics, Ben-Gurion University of the Negev, P.O.B. 653, Beer-Sheva 84105, Israel}

\date[]{Physica Scripta {\bf T151}, 014035 (2012). Special issue. Proceedings of FQMT conference (Prague, 2011).}

\begin{abstract}
The analysis of the response to driving in the case of weakly chaotic or weakly interacting systems 
should go beyond linear response theory. Due to the ``sparsity" of the perturbation matrix, 
a resistor network picture of transitions between energy levels is essential. 
The Kubo formula is modified, replacing the "algebraic" average over the squared 
matrix elements by a ``resistor network" average. Consequently the response becomes 
semi-linear rather than linear. Some novel results have been obtained in the context 
of two prototype problems: the heating rate of particles in Billiards with vibrating walls;  
and the Ohmic Joule conductance of mesoscopic rings driven by electromotive force.      
Respectively, the obtained results are contrasted with the ``Wall formula" and the ``Drude formula".  
\end{abstract}



\maketitle

\section{Introduction}

\begin{figure}
\centering

\includegraphics[width=0.5\hsize]{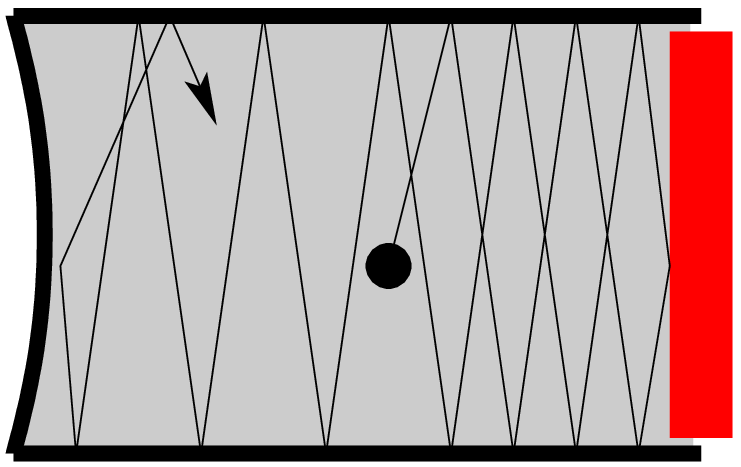} 

\vspace*{2mm}

\includegraphics[width=0.45\hsize]{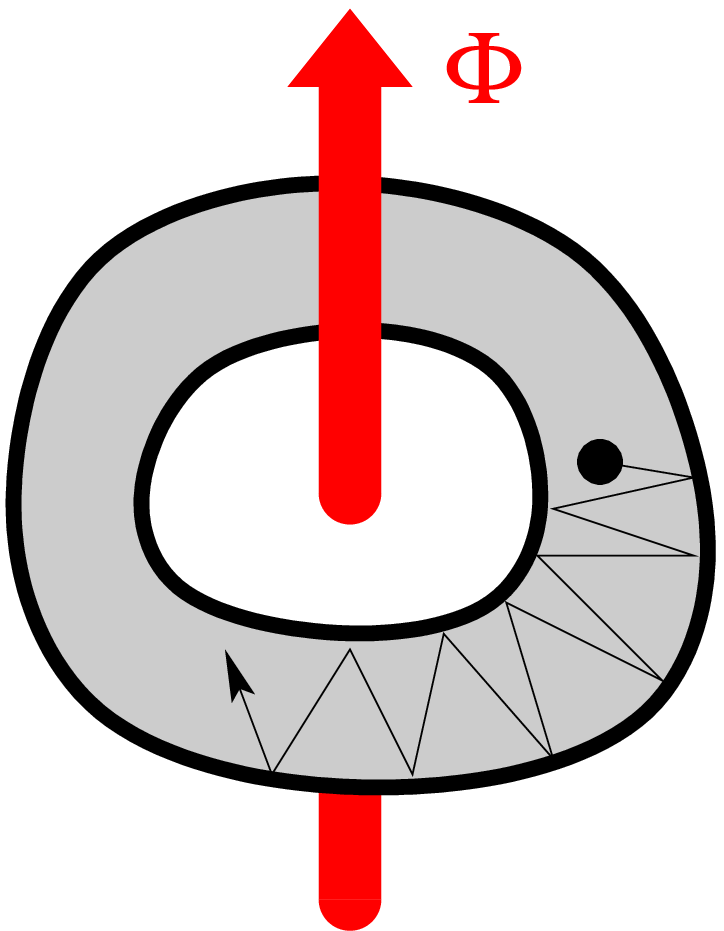}

\caption{
Model systems: a Billiard with a moving wall (upper panel), 
and a Ring with a time dependent magnetic flux (lower panel).
The deviation of the Billiard from integrability
is quantified by a parameter $u$. It is due to the deformation 
of the boundary (as in the figure) or due to a deformation 
of the potential floor (not illustrated). 
In the lower panel the Ring is regarded 
as a rectangular-like billiard with periodic boundary 
conditions in one of its coordinates. In the numerics 
the Ring has been modeled as a tight binding array 
of dimensions ${L\times \mathcal{M}}$. In the latter case 
the non-integrability was due to on-site disorder~$W$.  
} 

\label{f1}
\end{figure}

\begin{figure}
\centering

\ \ \ \includegraphics[clip,width=0.65\hsize]{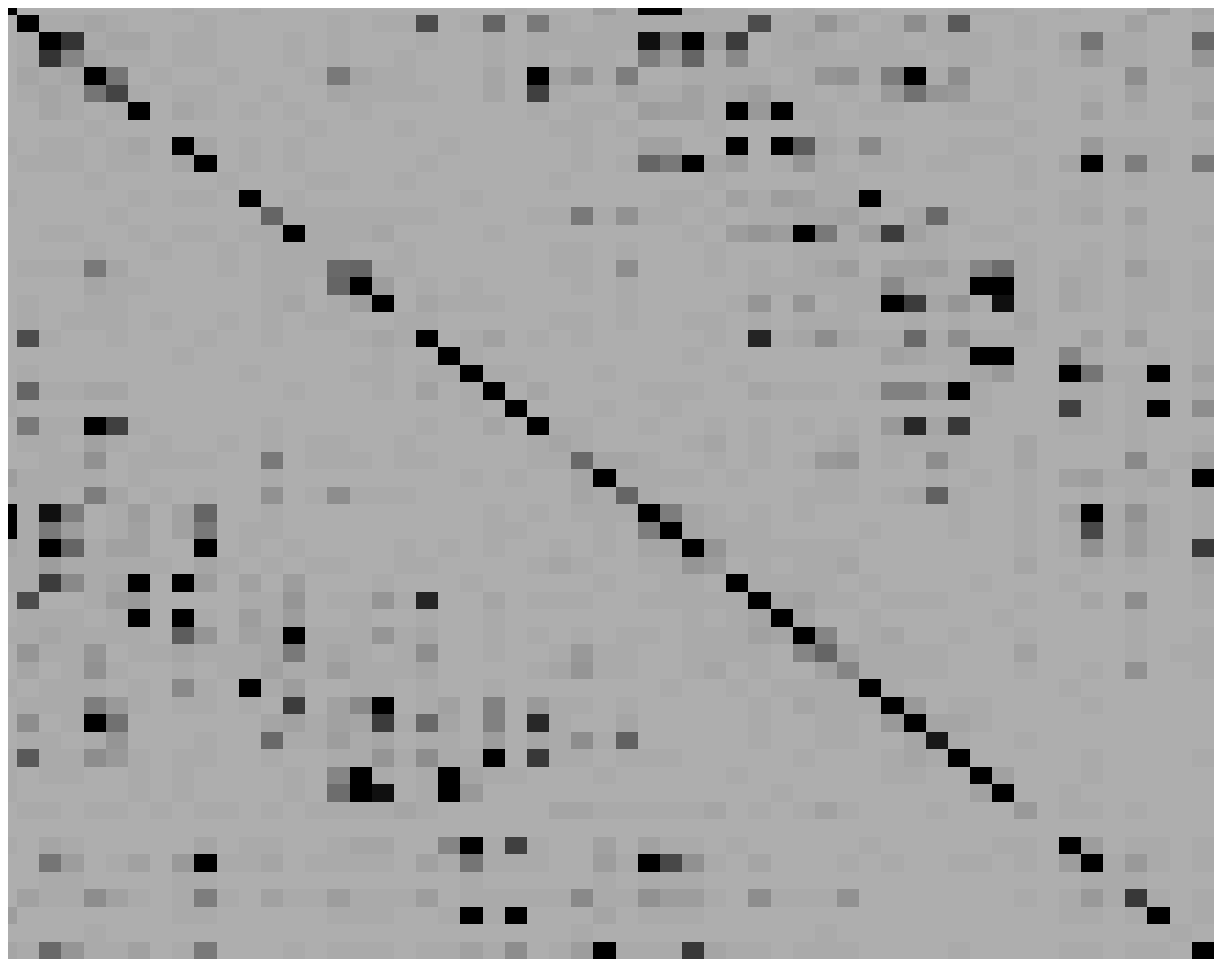}

\includegraphics[clip,width=0.85\hsize]{cqc_F_alg_med_vs_omega_deform_TLK} 

\caption{
The upper panel is an image of the matrix ${\bm{X}}$ 
for a billiard very similar in shape to that of \Fig{f1}. 
This matrix is "sparse". This can be deduced either by inspection,  
or by looking in the lower panel where the {\em average} 
value $\langle x \rangle$ of the elements (upper curves) 
is calculated as a function of the distance $\omega$ 
from the diagonal, and contrasted with the much smaller {\em median} values (lower curves). 
For further details see \mycite{cqb,cqc}. 
} 

\label{f2}
\end{figure}

\begin{figure}
\centering

\includegraphics[clip,width=0.9\hsize]{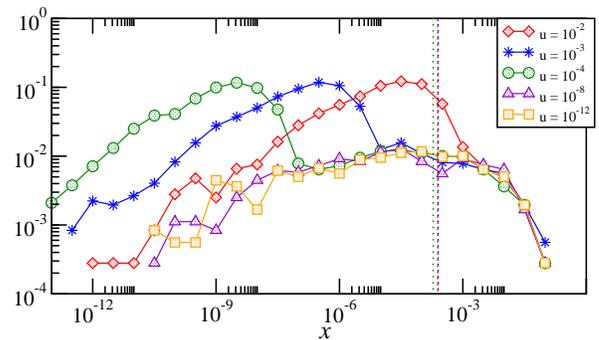}

\caption{
Histogram of the in-band elements $x$ of the matrix ${\bm{X}}$
for a rectangular billiard that has deformed 
potential floor. Different symbols refer 
to different values of the deformation parameter~$u$.
The vertical line is the {\em average} value 
of the elements, while the {\em median} 
is roughly at the locations of the peaks. 
The former unlike the latter is not sensitive 
to the degree of deformation.   
For details see \mycite{kbw}.
} 

\label{f3}
\end{figure}

This presentation concerns driven systems, like those 
illustrated in \Fig{f1}, whose dynamics is generated
by an Hamiltonian that is represented by a matrix 
that has the generic structure  
\be{1}
\mathcal{H}_{\tbox{total}} \ \ = \ \ \mbox{diag}\{ E_n \} -  f(t) \{ V_{nm} \}  
\eeq
Here $E_n$ are the ordered energy levels of the unperturbed system. 
Their density $\varrho$[states/energy] is assumed to be roughly uniform.  
The system is driven by a low frequency stationary driving source~$f(t)$.
The elements of the perturbation matrix are $V_{nm}$. 
The induced transitions have rates that are proportional to $|V_{nm}|^2$.
We define 
\be{2}
\bm{X} \ \ =  \ \  \{ |V_{nm}|^2 \}   
\eeq
Our interest concerns Hamiltonians that have a ``sparse" perturbation 
matrix. This means that the majority of elements in $\bm{X}$ are
small. To be more precise we assume that these elements 
have a log-wide distribution with a {\em median} that is much smaller 
compared with the {\em average}. An example for such matrix is 
given in \Fig{f2}, and a typical histogram of the elements is 
presented in \Fig{f3}.

The question that we ask is simple: Given $\bm{X}$, 
what is the calculation that should be done in 
order to get the energy absorption rate (EAR).
It makes sense that the result should be proportional 
to some weighted average $\langle\langle \bm{X} \rangle\rangle$ 
over the matrix elements.  Indeed, within the 
framework of linear response theory (LRT) the Kubo formula 
is doing just that - a weighted {\em algebraic} average.

One should realize that the use of the Kubo formula 
for EAR calculation can be justified only in the very 
weak driving limit, provided there is a background ``bath" 
that maintains quasi-equilibrium at any moment. 
However, if the driving is not very weak, compared 
with the relaxation, then one should be worried:
in order for the system to heat up, it is essential 
to have {\em connected} sequences of transitions, 
else the system is ``stuck". There is an obvious analogy 
here with a resistor network calculation: due to the 
sparsity the energy absorption somewhat resembles 
a percolation process.      
 
The bottom line of the above considerations is 
that the {\em algebraic} average of Kubo $\langle\langle \bm{X} \rangle\rangle_a$,
should be replaced by a resistor network average $\langle\langle \bm{X} \rangle\rangle_s$, 
whose value is much smaller if the matrix is ``sparse".
This should be regarded as an {\em anomaly} in the 
theory of response:  it is an effect that arises 
upon quantization. Namely, one can characterize the sparsity 
of $\bm{X}$ by a parameter ${0<s<1}$ that is absent   
in the classical context, but has a dramatic effect 
in the quantum analysis.

A few words are in order regarding the literature. We go here 
beyond the conventional random matrix theory (RMT) perspective 
of \mycite{wigner,bohigas}, because we are dealing with ``sparse" 
matrices \mycite{sparse1,sparse2,sparse3,sparse4}, 
possibly banded \mycite{mario1,mario2,mario3,mario4,prosen1,prosen2}.
Our view of LRT follows that of \mycite{ott1,ott2,ott3,jar1,jar2,wilk,WA,robbins}.
In the context of Billiards, LRT implies 
the ``Wall formula" \mycite{wall1,wall2,wall3,frc,dil,wlf},
while in the context of mesoscopic conductance  
LRT implies the ``Drude formula"  \mycite{kamenev}.
In both cases one should take into account {\em corrections}  
that are related to correlations and level statistics.   
The quest for {\em anomalies} that cannot be explained 
by introducing corrections within the framework of LRT, 
but go beyond LRT, has some history \mycite{wilk,WA,crs,rsp,krav1,krav2,lc}.
The line of study regarding the anomaly that 
arises due to ``sparsity" is documented 
in \mycite{kbr,bls,slr,bld,kbd,kbw,cqb,cqc}, 
see acknowledgment. 
The resistor network analysis that is introduced below 
is inspired by \mycite{mott,miller,AHL,pollak,VRHbook,kbv}, 
but generalizes its scope in a somewhat revolutionary way.  \\

{\em Outline.-- } We first introduce with more details 
the model systems of \Fig{f1}. Then we outline the 
formalism of the EAR calculation, that is based on 
a simple Fermi golden rule (FGR) picture. Finally 
we present results that we have obtained for the 
dependence of the absorption coefficient or the conductance 
on the sparsity, where the latter is controlled by 
the degree of deformation or by the disorder in the system. 
Two appendices gives extra details on the resistor network 
calculation and on what we call ``resistor network average".

\section{The model systems}

Assume that we have $N$ non-interacting particles 
in a ``box". For presentation purpose assume 
a rectangular-like two dimensional billiard shaped box 
as in \Fig{f1}a, or optionally, imposing periodic boundary 
conditions in once direction, a ring shaped box
as in \Fig{f1}b. The box is slightly deformed:
either its walls are slightly curved, or optionally 
the potential floor is not flat, e.g. due to some 
scatterer or disorder. The system is driven by 
a low frequency stationary source. In \Fig{f1}a 
the driving is induced by moving a ``piston", 
while in \Fig{f1}b it is by varying a magnetic flux 
through the ring. The Hamiltonian matrix in the
unperturbed energy basis takes that form of \Eq{e1}.
In the case of the driven ring we have the identifications 
\be{3}
f(t) & \ \mapsto \ &  \Phi(t)  \\
V_{nm} & \ \mapsto \ &  (e/L) \, v_{nm}  
\eeq
where $\Phi(t)$ is the magnetic flux, and $v_{nm}$ 
are the matrix elements of the velocity operator.
Note that by Faraday law $-\dot{\Phi}$ is the 
electromotive force (EMF).

The driving induces transitions between energy levels.
We assume stationary driving source, and define 
the its power spectrum as 
\be{5}
{\tilde S(\omega)}  \ \ = \ \ \mbox{FT}\  {\langle \dot{f}(t)\dot{f}(0)\rangle}
\eeq
Note that as far as EAR is concerned, 
it is a non-zero $\dot{f}$  that makes 
the Hamiltonian time dependent, 
corresponding to the EMF in the case of the ring. 
We assume low frequency driving. Accordingly we write 
\be{6}
\tilde{S}(\omega) \ \ = \ \ 2\pi \overline{|\dot{f}|^2} \ \delta_{c}(\omega)
\eeq
where the prefactor is the RMS value that characterizes 
the driving intensity, and $\delta_{c}(\omega)$ is a broadened delta function 
whose line shape reflects the spectral content of the driving.

\section{The energy absorption rate}

The driving induce transitions between energy levels, 
which implies diffusion in energy space. This diffusion
is characterized by a coefficient $D$[energy$^2$/time] 
for which we would like to have a formula. 
Assuming that $D$ is known, the EAR is given by the 
following expression:
\be{7}
\mbox{EAR} \ \ = \ \ \mbox{density} \times D 
\eeq
This is a straightforward generalization of Einstein type 
relations that are discussed in \mycite{wilk} and in greater 
details in \mycite{frc}. We can call it a diffusion-dissipation relation.  
What we label in \Eq{e7} as ``density" stands for 
the number of particles ($N$) per energy, 
meaning $N/T$ in the case of a Boltzmann occupation at tempeature~$T$, 
or $\varrho$ in the case of a low temperature Fermi occupation. 
In the latter case the role of the temperature 
is overtaken by the Fermi energy, namely~${\varrho \sim N/E_{\tbox{F}}}$.  

What we would like to have, is a theory that allows 
the calculation of $D$. The formulas that we would 
like to advertise is 
\be{8}
D \ \ = \ \ \pi \varrho
\ \langle\langle | {V_{nm}}|^2 \rangle\rangle \ \overline{\dot{f}^2}
\eeq
Depending on the interpretation of $\langle\langle | {V_{nm}}|^2 \rangle\rangle$
this is the Kubo formula of LRT, or its resistor-network variation.
In the latter case we refer to the results as the outcome 
of semi-linear response theory (SLRT).
The latter term indicates that $\langle\langle | {V_{nm}}|^2 \rangle\rangle_s$
unlike $\langle\langle | {V_{nm}}|^2 \rangle\rangle_a$ is a semi-linear 
rather than linear operation. 
The derivations of both the LRT and the SLRT variations 
of \Eq{e8} are outlined in the next sections. 

In the case of an EMF driven Ring, 
it is convenient to re-write 
the EAR formulas \Eq{e7} as 
\be{9}
\mbox{EAR} \ \ = \ \ G \, \overline{\dot{f}^2}
\eeq
where the so-called mesoscopic conductance 
is given by the expression
\be{10}
G \ \ = \ \ \pi \varrho^2  
\ \left(\frac{e}{L}\right)^2 
\ \langle\langle |v_{mn}|^2 \rangle\rangle
\eeq
One can regard $G$ as a mesoscopic version of 
the absorption coefficient, while \Eq{e9} can be regarded 
as the mesoscopic version of Joule law.

\section{The Fermi golden rule picture and the Kubo formula}

The Hamiltonian in the standard basis is \Eq{e1}.
We can transform it to the adiabatic basis: 
\be{11}
\tilde{\mathcal{H}} \ \ = \ \ \mbox{diag}\{ E_n \} -  {\dot{f}(t)} \left\{ \frac{i {V_{nm}}}{E_n-E_m} \right\}
\eeq
The FGR transition rate from $E_m$ to $E_n$ 
due to the low frequency noisy driving is:
\be{12}
w_{nm} \ \ = \ \ 
\left|\frac{ {V_{nm}}}{E_n-E_m}\right|^2 \ \tilde{S}(E_n{-}E_m)
\eeq
The FGR transitions lead to diffusion in energy space.
Assuming that there is a background relaxation 
process that maintains at any moment quasi-equilibrium 
with occupation probabilities $p_n$ that are the same 
as in the absence of driving, we get for the 
driving induced diffusion  
\be{13}
D \ \ = \ \ 
\sum_n p_n \left[\frac{1}{2} \sum_m (E_m{-}E_n)^2 \  w_{mn} \right]
\eeq
leading to the Kubo formula 
\be{14}
D \ \ = \ \ \int_{0}^{\infty}  {\tilde C(\omega)}  {\tilde S(\omega)} \frac{d\omega}{2\pi} 
\eeq
with the spectral function  
\be{15}
\tilde{C}(\omega) &=& \mbox{FT}\  {\langle V(t)V(0)\rangle}
\\ \nonumber
&=& \sum_n p_n \sum_{m}|V_{nm}|^2\ 2\pi\delta(\omega - (E_m{-}E_n))
\eeq
Note that this spectral function reflects 
the band profile of $\bm{X}$, as defined in \Eq{e2}, 
and illustrated in \Fig{f2}.

It is important to realize that the Kubo formula \Eq{e14} 
is a linear functional of $\tilde{S}(\omega)$.
The dependence on the matrix elements of $\bm{X}$ 
is linear too: 
\be{16}
D \ \ = \ \ 
\left[
\pi\sum_{n,m} p_n |V_{mn}|^2  \ \delta_{c}(E_m{-}E_n)
\right]
\ \overline{\dot{f}^2} 
\eeq
This can be formally written as \Eq{e8} 
with an implied definition of the weighted algebraic 
average $\langle\langle | {V_{nm}}|^2 \rangle\rangle_a$.

For completeness we note that in the context 
of conductance calculation the popular textbook version 
of the Kubo formula is  
\be{17}
G = \pi  \left(\frac{e}{L}\right)^2  
\sum_{n,m} |v_{mn}|^2  
\ \delta_{{T}}(E_n{-}E_F) 
\ \delta_{c}(E_m{-}E_n)
\eeq
This expression is implied by \Eq{e16}, 
with averaging over the levels in the vicinity 
of the Fermi energy. Namely ${p_n =\varrho^{-1}\delta_{T}(E_n{-}E_F)}$, 
where the width of $\delta_T()$ is determined by the temperature.

\begin{figure}
\includegraphics[clip,width=0.8\hsize]{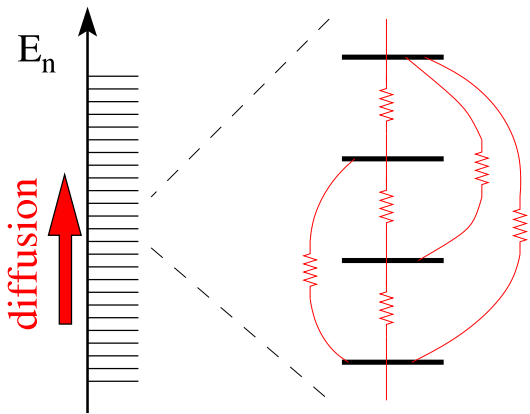}

\caption{
The driving induces transitions between levels $E_n$ of a closed system, 
leading to diffusion in energy space, and hence an associated heating. 
The diffusion coefficient $D$ can be calculated using a resistor network analogy. 
Connected sequences of transitions are essential in order 
to have a non-vanishing result, as in the theory of percolation.
}

\label{f4}
\end{figure}

\section{The calculation of the diffusion coefficient}

We would like to consider circumstances in which 
the driving induced transitions are faster compared 
with the background relaxation. In such circumstances  
the occupation probabilities~$p_n$ are no longer 
as in equilibrium. For the purpose of analysis we 
simply neglect the bath, and describe the dynamics 
by a rate equation
\beq
\frac{dp_n}{dt} = - \sum_m  w_{nm} (p_n-p_m)   
\eeq
where the rates $w_{nm}$ are determined by \Eq{e12}.   
The matrix ${\bm{w}=\{w_{nm}\} }$ can be regarded as a quasi 
one-dimensional network. See \Fig{f4}. 
Optionally one may interpret the rate equation as 
a probabilistic description of a random walk process, 
where $w_{nm}$ is the probability 
to hop from $m$ to $n$ per unit time.
The local spreading is described by 
\beq
\mbox{Var}(n) \ \ &=& \  \ \sum_{n} [w_{n,n_0}t] \, (n-n_0)^2 \\ 
\ \ &\equiv& \  \ 2\tilde{D}_{\tbox{local}} t
\eeq
It follows that the course-grained spreading 
should be described by a diffusion equations 
\beq
\frac{\partial p_n}{\partial t} 
\ \  = \ \  \tilde{D}\frac{\partial^2}{\partial n^2}p_n  
\eeq
where $n$ is regarded as a continuous variable.
The diffusion equation is formally a continuity equation 
\beq
\frac{\partial p_n}{\partial t} 
\ \ = \ \  -\frac{\partial }{\partial n} I_n  
\eeq
where the current is given by Fick's law:
\beq
I_n \ \ = \ \ -\tilde{D} \frac{\partial }{\partial n} p_n
\eeq
If we have a sample of length $N$ then   
\beq
I \ \ = \ \ -\frac{\tilde{D}}{N} \times [p_N-p_0]
\eeq
This shows that $\tilde{D}/N$ is formally like 
the inverse resistance of the chain:
it is the ratio between the current and 
the ``potential difference". 
We therefore can use standard recipes 
of electrical engineering in order to 
calculate its value. For example, if we have only 
near-neighbor transitions then 
``adding connectors in series" implies  
\beq
\frac{\tilde{D}}{N} \ \ = \ \ \left[\sum_{n=1}^N \frac{1}{w_{n,n{-}1}}\right]^{-1} 
\eeq
In general we use the notation $\tilde{D}=[[ \bm{w} ]]$, where 
the doubled brackets stand for inverse resistivity calculation,   
as discussed in App.\ref{a1}. 
   
In the above analysis we have assumed unit distance between sites.
If the mean level spacing is $\varrho^{-1}$, the expression 
for the diffusion coefficient should be re-scaled as follows:
\beq
D \ \ = \ \  \varrho^{-2} [[ \bm{w} ]] 
\eeq
Recall that the $w_{nm}$ are given by \Eq{e12}, 
it follows that 
\beq
D = \varrho^{-2} \left[\left[ \left|\frac{ {V_{nm}}}{E_n-E_m}\right|^2 \ 2\pi \overline{|\dot{f}|^2} \delta_{c}(E_n{-}E_m) \right]\right]
\eeq
leading to \Eq{e10}, with an implied definition 
of the resistor network average $\langle \langle |V_{nm}|^2 \rangle \rangle_s$.  
The definition and the calculation of the latter  
are further discussed in App.\ref{a2}.

\section{The wall formula and beyond}

The roughest estimate for the diffusion that is induced 
by a vibrating wall is known as the ``Wall formula" \mycite{wall1,wall2,wall3,frc,dil,wlf}. 
Its original derivation is based on a simple kinetic picture: 
it is based on the assumption that collisions with the vibrating wall 
are not correlated. This leads in the two dimensional case \mycite{frc,cqc}
to the result 
\beq
D_0 \ \ = \ \ \frac{4}{3\pi} \frac{\mass^2 v_{\tbox{E}}^3}{L_x}  \  {\overline{\dot{f}^2}} 
\eeq
where $\mass$ is the mass of the particles, and $L_x$ is 
the linear dimension of the box as illustrated in \Fig{f1}.
The result assumes a microcanonical preparation at 
energy~$E$, and we have defined $v_{\tbox{E}}=[2E/\mass]^{1/2}$.
If we have a Boltzmann occupation, the expression 
should be averaged accordingly. If we have low 
temperature Fermi occupation, what counts in \Eq{e7} 
is the the value at $E=E_{\tbox{F}}$.     

Within the framework of LRT the same result is obtained 
from the Kubo formula \Eq{e14}, provided $\tilde{C}(\omega)$ is flat, 
i.e. provided there are no correlations between collisions. 
In practice there are correlations 
leading to ${D = g_c D_0}$ with $g_c$ that can be either 
smaller or larger than unity depending on the geometry.
In the quantum calculation $g_c$ is slightly affected   
by the level spacing statistics. 

Within the framework of SLRT one has to calculate the 
resistor network average of the $\bm{X}$ matrix. 
If this matrix is sparse, the result becomes very suppressed, 
leading to ${D = g_s g_c D_0}$, where 
\beq
g_{s} 
\ \ \equiv \ \ 
\frac{\langle\langle |V_{nm}|^2 \rangle\rangle_{s}}
{\langle\langle |V_{nm}|^2 \rangle\rangle_{a}}
\eeq
We emphasize that $g_s$ reflects an anomaly: 
it depends on the sparsity $s$ of the matrix,  
a parameter that has no 
meaning in the classical context.  
For more details, including RMT analysis 
of this dependence see \mycite{kbd}.

\begin{figure}
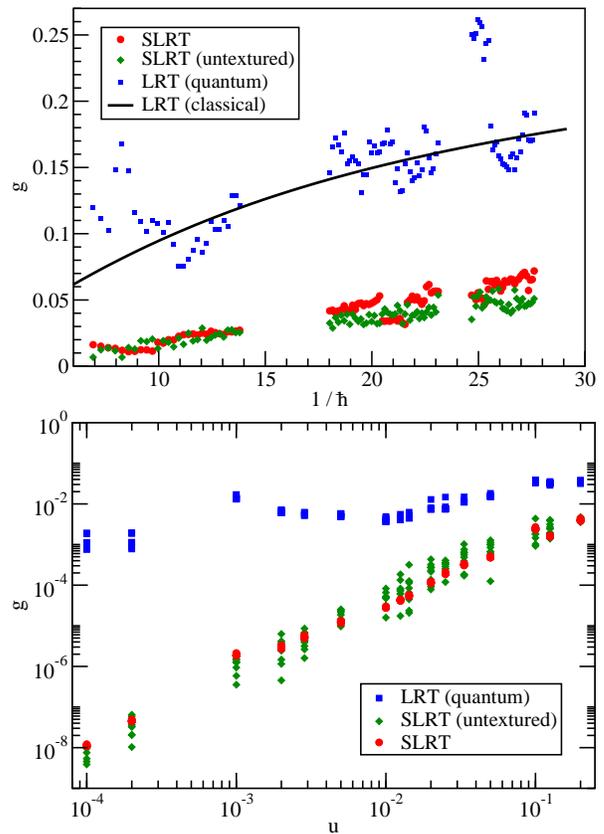

\centering
\includegraphics[clip,width=0.9\hsize]{cqb_g_vs_h}

\includegraphics[clip,width=0.9\hsize]{cqc_g_vs_u}

\caption{
The scaled absorption coefficient $g_c$ (LRT) 
and $g=g_sg_c$ (SLRT) versus the dimensionless $1/\hbar$ (upper panel),
and versus the dimensionless deformation parameter $u$ (lower panel).
The input for this analysis is the matrix $\bm{X}$ of \Fig{f2}. 
The calculation of each point has been carried out on a $100 \times 100$ 
sub-matrix of $\bm{X}$  centered around the $\hbar$ implied energy~$E$. 
The ``untextured'' data points are calculated for an artificial  
random matrices with the same bandprofile and sparsity.  
The complementary lower panel is oriented to show the small~$u$ 
dependence within an energy window that corresponds to ${1/\hbar\sim9}$. 
For further details see \mycite{cqb,cqc}. 
}

\label{f5}
\end{figure}

Some numerical results for $g_c$ (LRT) 
and $g=g_sg_c$ (SLRT) are presented in \Fig{f5}.
The calculation is done with 
the $\tilde{C}(\omega)$ of \Fig{f2}.
The $\hbar$ dependence of the LRT ``classical" result 
is due to the lower cutoff of the $d\omega$ integral 
in the Kubo formula \Eq{e14}, 
that is sensitive to mean level spacing. 
If $\tilde{C}(\omega)$ were ``flat" the result 
would not be much sensitive to~$\hbar$. 
As expected the SLRT calculation gives a 
much smaller result. The effect becomes 
more conspicuous for smaller deformations, 
for which the sparsity is ``stronger".

\section{The Drude formula and beyond}

The EAR by particles that are driven by an oscillating electric field,  
due to an induced EMF, is very similar problem to that of particles 
driven by an oscillating wall. Here the simple result that 
is based on kinetic considerations is known as the ``Drude formula". 
As in the case of the ``Wall formula" it is assumed that scattering 
events are uncorrelated, leading to the estimate    
\beq
D_0 \ \ = \ \ \left(\frac{e}{L}\right)^2 \ v_{\tbox{E}} \ell \ \ \overline{\dot{\Phi}^2} 
\eeq
where $e$ is the charge of the electron, $L$ is the 
length of the ring, and $\ell$ is the mean 
free path between collisions.  
Note that $\mathcal{D}=v_{\tbox{E}} \ell$  
is the spatial diffusion coefficient, 
while $(e\dot{\Phi}/L)$ is the energy that 
is gained per circulation. 
The implied result for the 
conductance can be written as 
\beq
G_0 \ \ = \ \ \frac{e^2}{2 \pi \hbar} \ \mathcal{M} \ \frac{\ell}L
\eeq
where $\mathcal{M}=\mass v_{\tbox{E}}L_y/\pi$  
is the number of open modes. This way of writing the Drude formula 
is very illuminating because it reflects Ohm law,  
and the units are as in the Landauer formula. 
For clarity we have restored the $\hbar$ in this formula.

Within the framework of LRT, the same result is obtained 
from the Kubo formula \Eq{e14}, provided $\tilde{C}(\omega)$   
of \Eq{e15} is a Lorentzian that reflects an exponential decay of 
the velcoity-velocity correlation function. 
In practice there are extra correlations 
leading to ${D = g_c D_0}$ with $g_c$ that can be either 
smaller or larger than unity depending on the geometry.
In the quantum calculation $g_c$ is slightly affected   
by the level spacing statistics, and the correction 
is of order $(\varrho\omega_c)^{-1}$.  This is sometimes 
regarded as a variation of weak localization corrections \mycite{kamenev}.

Within the framework of SLRT one has to calculate the 
resistor network average of the matrix $\{|v_{nm}|^2\}$.
Here one should distinguish between different regimes, 
depending on the strength~$W$ of the disorder. 
In the Born approximation the mean free path is ${\ell \propto W^2}$, 
while the localization length is ${\ell_{\infty} \approx \mathcal{M}\ell}$.
The diffusive regime, where there is no issue of sparsity, 
requires an intermediate strength of disorder, 
such that ${\ell \ll L \ll \ell_{\infty}}$.   
For either stronger or weaker disorder, 
the matrix $\{|v_{nm}|^2\}$ becomes ``sparse".  
This is because the eigenstates become non-ergodic: 
either they are localized in real space (for strong disorder) 
or in mode space (for weak disorder). 
Note also that for very weak disorder ("clean" ring) 
each eigenstate occupies a single mode (up to small correction).   
For detailed analysis that supports the above picture see \mycite{kbd}.

\begin{figure}

\includegraphics[clip, width=\hsize]{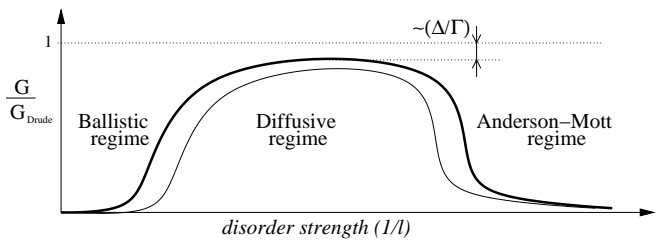}

\caption{
Schematic illustration that sketch the dependence 
of the DC mesoscopic conductance on the strength of the disorder.
It should be regarded as a caricature of \Fig{f7}. 
The level width $\Gamma=\hbar\omega_c$ affects the so-called 
weak localization correction in the diffusive regime.
In the other two regimes of either weak or strong disorder, 
the perturbation matrix becomes ``sparse" and 
consequently $G$ is suppressed compared with Drude.  
}

\label{f6}
\end{figure}

\begin{figure}
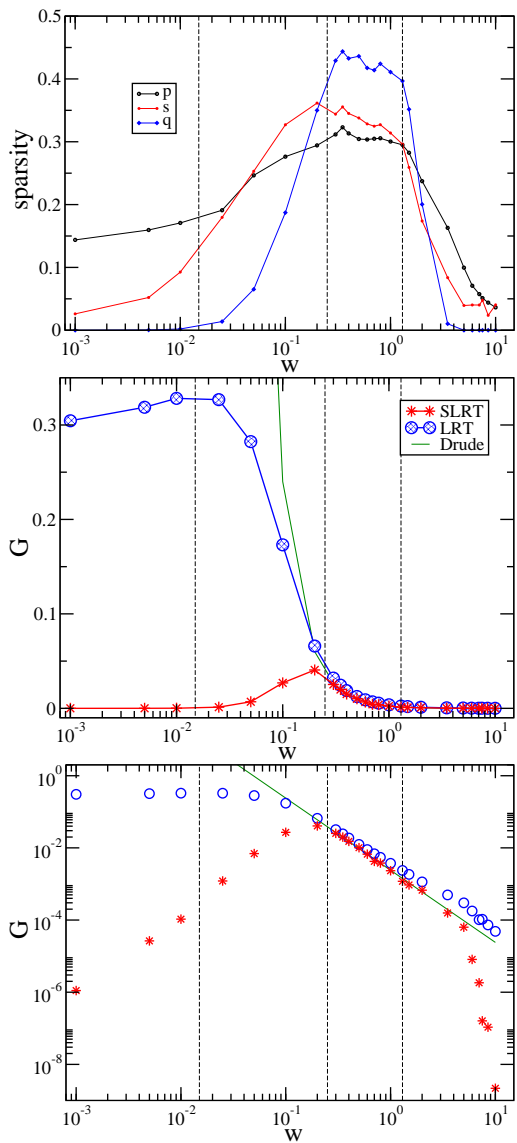


\includegraphics[clip, width=0.78\hsize]{spq_vs_w_normal}
\includegraphics[clip, width=0.78\hsize]{G_vs_W_normal}
\includegraphics[clip, width=0.78\hsize]{G_vs_W_log}

\caption{
Various measures of sparsity (upper panel), 
and the mesoscopic conductance (lower panels) 
as a function of the disorder~$W$ in the Ring.
The vertical line separate between  
the clean, ballistic, diffusive and localization regimes.
Note that the scaled conductance in arbitrary units 
equals $\langle\langle |v_{mn}|^2 \rangle\rangle$.
The Drude, the LRT and the SLRT results 
are displayed in both normal and log scale.
We see that in the ballistic regime the SLRT conductance 
becomes worse as the disorder becomes weaker, 
in opposition with the Drude expectation.
For further details see \mycite{bld,kbd}
}

\label{f7}
\end{figure}

The expectation with regard o the dependence of $G/G_0$ 
on the strength of the disorder is sketched in \Fig{f6}.  
Some numerical results for both the LRT and the SLRT conductance 
are presented in \Fig{f7}. The calculation is done 
for a tight binding model. The SLRT result in the Anderson 
localization regime is completely analogous to the reasoning 
of variable range hopping \mycite{mott,miller,AHL,pollak,VRHbook}, 
as explained in \mycite{kbv}. It should be appreciated 
that in our approach all regimes are treated on equal footing. 

In the ballistic regime, contrary to the Drude expectation, 
the conductance becomes worse as the disorder is reduced.
This looks strange, but can easily be rationalized 
if we think about the extreme case of no disorder: 
in the absence of scattering the particle stays all the 
time in the same mode; hence irreversible diffusive 
spreading in energy is impossible.

\section{Conclusions}
\label{sec:Conclusions}

The random matrix approach of Wigner ($\sim 1955$) is based  
on the observation that in generic circumstances 
the perturbation can be represented by a random matrix 
whose elements are taken from a Gaussian distribution.
Our interest in this presentation concerned a restricted 
class of ``sparse" systems for which this observation 
does not hold. In such {\em weak} quantum chaos circumstances 
the elements are characterized by a log-wide distribution. 
Consequently, the response, and in particular the energy absorption,  
are similar to a percolation process, and their analysis 
requires a novel resistor-network approach. 

Besides the quantitative issue, the experimental fingerprint 
of the resistor-network calculation 
is the implied semi-linearity of the response. 
In the SLRT regime, i.e. if the driving is the 
predominant mechanism for transitions between levels,  
one expects the combined effect of two independent 
sources to be super-linear, namely, 
\beq
D\big[\tilde{S}_a(\omega) + \tilde{S}_b(\omega)\big]  \ \ > \ \ D\big[\tilde{S}_a(\omega)\big] + D\big[\tilde{S}_b(\omega)\big]
\eeq
but still semi-linear $D\big[ \lambda \tilde{S}(\omega)\big] = \lambda \ D\big[\tilde{S}(\omega)\big]$. 
We have provides in this presentation two prototype 
examples where an SLRT anomaly can arise: heating of 
particles that are trapped in billiards with vibrating walls;  
and Joule heating of charged carriers that are driven by an 
induced electro-motive force.


\clearpage
\appendix

\section{The resistor network calculation}
\label{a1}

In this appendix we explain how the inverse 
resistivity $G=[[\bm{G}]]$ of a one-dimensional 
resistor network ${\bm{G}\equiv\{G_{nm}\}}$ is calculated. 
We use the language of electrical engineering for this purpose.
In general this relation is semi-liner rather than linear, 
namely  ${[[\lambda \bm{G}]]=\lambda[[\bm{G}]]}$, 
but ${[[\bm{A}+\bm{B}]] \ne [[\bm{A}]]+[[\bm{B}]]}$.

There are a few cases where an analytical expression 
is available. 
If only near neighbor nodes are connected, 
allowing ${G_{n,n+1}=g_n}$ to be 
different from each other, 
then ``addition in series" implies 
that the inverse resistivity calculated 
for a chain of length~$N$ is  
\be{0}
G \ \ = \ \ \left[\frac{1}{N}\sum_{n=1}^{N} \frac{1}{g_n}\right]^{-1}
\eeq
If $G_{nm}=g_{n-m}$ is a function 
of the distance between the nodes $n$ and $m$ 
then it is a nice exercise to prove 
that ``addition in parallel" implies 
\be{92}
G \ \ = \ \ \sum_{r=1}^{\infty} r^2 g_r
\eeq

In general an analytical formula for~$G$
is not available, and we have to apply 
a numerical procedure. For this purpose 
we imagine that each node~$n$ is connected 
to a current source $I_n$. The Kirchhoff    
equations for the voltages are    
\be{0}
\sum_m G_{mn} (V_n-V_m)  \  = \ I_n
\eeq
This set of equation can be written in a matrix form:
\be{0}
\tilde{\bm{G}} \bm{V}  \ = \ \bm{I}
\eeq
where the so-called discrete Laplacian matrix of the network is defined as 
\be{0}
\tilde{G}_{nm} =  \left[\sum_{n'}  G_{n'n}\right]\delta_{n,m} - G_{nm}
\eeq
This matrix has an eigenvalue zero which is associated  
with a uniform voltage eigenvector. Therefore, it has 
a pseudo-inverse rather than an inverse, and the Kirchhoff 
equation has a solution if and only if ${\sum_n I_n=0}$.      
In order to find the resistance between nodes ${n_{\tbox{in}}=0}$ 
and ${n_{\tbox{our}}=N}$, 
we set ${I_0=1}$ and ${I_N=-1}$ and ${I_n=0}$ otherwise,  
and solve for $V_0$ and $V_N$. 
The inverse resistivity is ${G=[(V_0-V_N)/N]^{-1}}$.

\section{The resistor-newtwork average}
\label{a2}

We use the notation $\langle\langle \bm{X} \rangle\rangle$
in order to indicate the weighted {\em average} value of its elements. 
First we would like to define the standard {\em algebraic} average. 
It is essential to introduce a {\em weight} function that defines 
the band of interest. In the physical context this function reflects 
the spectral content of the driving source. Namely, 
we define $F(r)$ as the normalized version of $\tilde{S}(\omega)$, 
such that $\sum_r F(r)=1$, where ${r=n-m}$ is the energy 
difference $\omega=E_n-E_m$ in integer units. 
The bandwidth in these dimensionless units (${b_c=\varrho\omega_c}$)       
it is assumed to be quantum mechanically large ($b_c \gg1$).
The algebraic average is defined in the standard way:
\be{0}
\langle\langle \bm{X} \rangle\rangle_a \ \ = \ \ 
\frac{1}{N} \sum_{n,m} F(n-m) \ X_{nm}
\eeq
where $N$ is the size of the matrix, 
which is assumed to be very large.  
The algebraic average is a {\em linear} operation, 
meaning that  
\be{0}
\langle\langle \lambda \bm{X} \rangle\rangle &=& \lambda \langle\langle \bm{X} \rangle\rangle
\\
\langle\langle \bm{X} + \bm{Y} \rangle\rangle &=& 
\langle\langle \bm{X} \rangle\rangle + \langle\langle \bm{Y} \rangle\rangle 
\eeq

There are different type of ``averages" in the literature, 
such as the {\em harmonic} average,  {\em geometric} average, 
and we can also include the {\em median} in the same list.
All these ``averages" are {\em semi-linear} operations because
only the  $\langle\langle \lambda \bm{X} \rangle\rangle = \lambda \langle\langle \bm{X} \rangle\rangle$ 
property is satisfied for them.   
Irrespective of the semi-linearity issue {\em any} type of 
average should satisfy the following requirement:  
if all the elements equal to the same number, 
then also the average should equal the same number.
 
In this presentation we highlight a new type of average 
that we call a {\em resistor-network} average:
\beq
\langle\langle \bm{X} \rangle\rangle_s \ \ \equiv \ \ 
\left[ \left[  2F(n{-}m) \, \frac{X_{nm}}{(n-m)^2} \right] \right]
\eeq
Writing the expression above as $[[ \tilde{w}_{nm} ]]$, 
one should realize that the $\tilde{w}_{nm}$ can be 
regarded as FGR transition rates.  
Using \Eq{e92} it is not difficult 
to show that if all the elements $X_{nm}$ are the same number, 
then also their resistor network average is the same number.
While in general 
\beq
\langle\langle \bm{X} \rangle\rangle_s \ \ < \ \ \langle\langle \bm{X} \rangle\rangle_a
\eeq
Typically the resistor network average is bounded 
from below by the median. In order to get a 
realistic estimate in the case of a ``sparse" matrix 
one can use a generalized variable range hopping 
scheme that we have developed in~\mycite{kbd}.


\begin{acknowledgments}
The current manuscript is based and reflects a line of study 
that has been carried out (in chronological order) in collaboration 
with \mycite{kbr,bls,slr,bld,kbd,kbw,cqb}: 
Tsampikos Kottos, Holger Schanz, Swarnali Bandopadhyay, Yoav Etzioni, 
Michael Wilkinson, Bernhard Mehlig, Alex Stotland, Rangga Budoyo, 
Tal Peer, Nir Davidson, and Louis Pecora.
The present contribution has been supported by 
the Israel Science Foundation (grant No.29/11)
\end{acknowledgments}


\clearpage

\begin{thebibliography}{10}



\bibitem{wigner} 
E. Wigner, Ann. Math {\bf 62} 548 (1955); {\bf 65} 203 (1957).

\bibitem{bohigas} 
O. Bohigas in 
{Chaos and quantum Physics}, 
Proc. Session LII of the Les-Houches Summer School, 
Edited by A. Voros and M-J Giannoni 
(Amsterdam: North Holland 1990).

\bibitem{sparse1} 
E.J. Austin, M. Wilkinson, 
{Europhys. Lett.} {\bf 20}, 589 (1992). 

\bibitem{sparse2} 
T. Prosen, M. Robnik, 
{J. Phys. A} {\bf 26}, 1105 (1993).

\bibitem{sparse3} 
Y. Alhassid, R.D. Levine, 
{Phys. Rev. Lett.} {\bf 57}, 2879 (1986).

\bibitem{sparse4}
Y.V. Fyodorov, O.A. Chubykalo, F.M. Izrailev, G. Casati, 
{Phys. Rev. Lett.} {\bf 76}, 1603 (1996).



\bibitem{mario1} 
M. Feingold, A. Peres, 
{\em Phys. Rev. A} {\bf 34} 591, (1986).

\bibitem{mario2} 
M. Feingold, D. Leitner, M. Wilkinson, 
{\em Phys. Rev. Lett.} {\bf 66}, 986 (1991). 
%
\bibitem{mario3} 
M. Wilkinson, M. Feingold, D. Leitner, 
{\em J. Phys. A} {\bf 24}, 175-182 (1991). 

\bibitem{mario4} 
M. Feingold, A. Gioletta, F.M. Izrailev, L. Molinari, 
{\em Phys. Rev. Lett.} {\bf 70}, 2936–2939 (1993).

\bibitem{prosen1} 
T. Prosen and M. Robnik, J. Phys. A 26 L319 (1993)

\bibitem{prosen2} 
T. Prosen, Ann. Phys. (N.Y.) 235, 115 (1994)


\bibitem{ott1}
E. Ott, 
{Phys. Rev. Lett.} {\bf 42}, 1628 (1979). 

\bibitem{ott2}
R. Brown, E. Ott, C. Grebogi, 
{Phys. Rev. Lett.} {\bf 59}, 1173 (1987).

\bibitem{ott3}
R. Brown, E. Ott, C. Grebogi, 
{J. Stat. Phys.} {\bf 49}, 511 (1987).

\bibitem{jar1}
C. Jarzynski, 
{Phys. Rev.} {\bf E 48}, 4340 (1993).

\bibitem{jar2}
C. Jarzynski, 
{Phys. Rev. Lett.} {\bf 74}, 2937 (1995).

\bibitem{wilk}
M. Wilkinson, 
{J. Phys. A} {\bf 21}, 4021 (1988).

\bibitem{WA}
M. Wilkinson, E.J. Austin, 
{J. Phys. A} {\bf 28}, 2277 (1995). 

\bibitem{robbins}
J.M. Robbins, M.V. Berry, 
{J. Phys. A} {\bf 25} L961 (1992). 



\bibitem{wall1}
D.H.E. Gross,
{Nucl. Phys. A} {\bf 240}, 472 (1975).

\bibitem{wall2}
J. Blocki, Y. Boneh, J.R. Nix, J. Randrup, M. Robel, A.J. Sierk, W.J. Swiatecki, 
{Ann. Phys.} {\bf 113}, 330 (1978).

\bibitem{wall3}
S.E. Koonin, R.L. Hatch, J. Randrup, 
{Nuc. Phys. A} {\bf 283}, 87 (1977). 

\bibitem{frc}
D. Cohen, 
{Annals of Physics} {\bf 283}, 175 (2000).

\bibitem{dil} 
A. Barnett, D. Cohen, E.J. Heller, 
{Phys. Rev. Lett.} {\bf 85}, 1412 (2000);  

\bibitem{wlf}
A. Barnett, D. Cohen, E.J. Heller, 
{J. Phys. A} {\bf 34}, 413 (2001). 




\bibitem{kamenev} 
For a review and further references see 
``(Almost) everything you always wanted to know about 
the conductance of mesoscopic systems" 
by A. Kamenev and Y. Gefen, Int. J. Mod. Phys. {\bf B9}, 751 (1995).



\bibitem{crs}
D. Cohen, 
{Phys. Rev. Lett.} {\bf 82}, 4951 (1999). 

\bibitem{rsp}
D. Cohen, T. Kottos, 
{Phys. Rev. Lett.} {\bf 85}, 4839 (2000).

\bibitem{krav1}
D.M. Basko, M.A. Skvortsov, V.E. Kravtsov, 
{Phys. Rev. Lett.} {\bf 90}, 096801 (2003).

\bibitem{krav2}
A. Silva, V.E. Kravtsov, 
{Phys. Rev. B} {\bf 76}, 165303 (2007).

\bibitem{lc}
T. Prosen, D.L. Shepelyansky, Eur. Phys. J. B {\bf 46}, 515 (2005).  




\bibitem{kbr}
D. Cohen, T. Kottos, H. Schanz, 
{J. Phys. A} {\bf 39}, 11755 (2006). 

\bibitem{bls}
S. Bandopadhyay, Y. Etzioni and D. Cohen, 
Europhysics Letters {\bf 76}, 739 (2006).

\bibitem{slr}
M. Wilkinson, B. Mehlig, D. Cohen, 
{Europhys. Lett.} {\bf 75}, 709 (2006).

\bibitem{bld} 
A. Stotland, R. Budoyo, T. Peer, T. Kottos, D. Cohen, 
{J. Phys. A} (FTC) {\bf 41}, 262001 (2008). 

\bibitem{kbd}
A. Stotland, T. Kottos, D. Cohen, 
{Phys. Rev. B} {\bf 81}, 115464 (2010).

\bibitem{kbw}
A. Stotland, D. Cohen, N. Davidson, 
{Europhys. Lett.} {\bf 86}, 10004 (2009).

\bibitem{cqb}
A. Stotland, L.M. Pecora and D. Cohen, 
Europhys. Lett. {\bf 92}, 20009 (2010). 

\bibitem{cqc}
A. Stotland, L.M. Pecora and D. Cohen, 
Phys. Rev. E {\bf 83}, 066216 (2011). 




\bibitem{mott} 
N.F. Mott, Phil. Mag. {\bf 22}, 7 (1970). 
\ N.F.~Mott and E.A.~Davis, 
Electronic processes in non-crystalline materials, 
(Clarendon Press, Oxford, 1971). 

\bibitem{miller}
A. Miller and E. Abrahams, Phys. Rev. {\bf 120}, 745 (1960).

\bibitem{AHL}
V. Ambegaokar, B. Halperin, J.S. Langer, 
Phys. Rev. B {\bf 4}, 2612 (1971). 

\bibitem{pollak}
M. Pollak, J. Non-Cryst. Solids {\bf 11}, 1 (1972).

\bibitem{VRHbook} 
B.I. Shklovskii and A.L. Efros, 
Electronic properties of doped semiconductors,
(Springer-Verlag Berlin Heidelberg 1984).

\bibitem{kbv}
D. Cohen, Phys. Rev. B 75, 125316 (2007). 


\end{thebibliography}
\end{document}